# Title

**Diff5T: Benchmarking Human Brain Diffusion MRI with an Extensive 5.0 Tesla K-Space and Spatial Dataset**

# Authors

Shanshan Wang#*[1,2], Shoujun Yu#[1,2], Jian Cheng#[3], Sen Jia[1], Changjun Tie[1,2], Jiayu Zhu[4], Haohao Peng[4], Yijing Dong[4], Jianzhong He[7], Fan Zhang[8], Yaowen Xing[4], Xiuqin Jia[6], Qi Yang[6], Qiyuan Tian[5], Hua Guo[5], Guobin Li[4], Hairong Zheng*[1,2]

# Affiliations

1. Paul C. Lauterbur Research Center for Biomedical Imaging, Shenzhen Institute of Advanced Technology, Chinese Academy of Sciences, Shenzhen, China
2. University of Chinese Academy of Sciences, Beijing, China
3. School of Computer Science and Engineering, Beihang University, Beijing, China
4. United-Imaging Healthcare, Shanghai, China
5. School of Biomedical Engineering, Tsinghua University, Beijing, China
6. Department of Radiology, Beijing Chaoyang Hospital, Capital Medical University Key Lab of Medical Engineering for Cardiovascular Disease, Ministry of Education, Beijing
7. Institute of Information Processing and Automation, College of Information Engineering, Zhejiang University of Technology
8. University of Electronic Science and Technology of China, Chengdu, China

Corresponding author(s): Shanshan Wang (ss.wang@siat.ac.cn); Hairong Zheng (hr.zheng@siat.ac.cn);

# Abstract

Diffusion magnetic resonance imaging (dMRI) provides critical insights into the microstructural and connectional organization of the human brain. However, the availability of high-field, open-access datasets that include raw k-space data for advanced research remains limited. To address this gap, we introduce Diff5T, a first comprehensive 5.0 Tesla diffusion MRI dataset focusing on the human brain. This dataset includes raw k-space data and reconstructed diffusion images, acquired using a variety of imaging protocols. Diff5T is designed to support the development and benchmarking of innovative methods in artifact correction, image reconstruction, image preprocessing, diffusion modelling and tractography. The dataset features a wide range of diffusion parameters, including multiple b-values and gradient directions, allowing extensive research applications in studying human brain microstructure and connectivity. With its emphasis on open accessibility and detailed benchmarks, Diff5T serves as a valuable resource for advancing human brain mapping research using diffusion MRI, fostering reproducibility, and enabling collaboration across the neuroscience and medical imaging communities.

# Background & Summary

The exceptional soft tissue contrast and versatility of magnetic resonance imaging (MRI) render it an indispensable diagnostic modality for various disorders, including neurological, musculoskeletal, and oncological diseases[1]. Diffusion MRI (dMRI)[2] is a unique and valuable modality for mapping brain tissue, microstructure and connectivity patterns from diffusion characteristics of endogenous water molecules.

Over the past few decades, the rapid advancement, availability, and diversity of large, open-source diffusion MRI datasets have significantly propelled human brain research. Notable datasets, such as the Human Connectome Project (HCP) [3–11], the Cambridge Centre for Aging and Neuroscience (Cam-CAN)[12], the UK Biobank[13,14], the Adolescent Brain Cognitive Development (ABCD) Study[7], the Amsterdam Ultra-high Field Adult Lifespan Database (AHEAD)[15,16], and the Amsterdam Open MRI Collection (AOMIC)[17], have provided critical resources for analyzing healthy and diseased human brain anatomical and connectional patterns in large scale. These datasets have enabled researchers to investigate white matter connectivity across the entire brain.

Despite these advancements, current open-access dMRI datasets were primarily acquired on 3.0T or 7.0T MRI systems, with their own. Especially, 3.0T dMRI, while offering shorter acquisition times, is constrained by lower spatial resolution and signal-to-noise ratio (SNR). Conversely, 7.0T dMRI has the potential to achieve superior spatial resolution and SNR but faces significant challenges, such as rapid transverse magnetization decay, increased B0 and B1+ inhomogeneity, and higher specific absorption rates (SAR)[3,18–20]. Therefore, it is important to investigate new ways for a better balance between spatial resolution, SNR, and acquisition time in multi-shell dMRI[21].

5.0T MRI systems are promising for bridging the gap between these two extremes. This intermediate field strength has already demonstrated application value in abdominal dMRI[21,22], cerebral vascular TOF-MRA[23] and cardiac imaging[24–26]. A 5.0T MRI system is expected to offer higher spatial resolution and SNR than 3.0T while reducing image artifacts compared to 7.0T[27]. The Diff5T dataset was created to provide the research community with a benchmark dataset supporting advancements in dMRI technology and applications.

In addition to images, Diff5T uniquely includes publicly accessible raw k-space data, allowing researchers to explore advanced reconstruction techniques. The complexity of the biophysical models used in dMRI often dictates the number of diffusion-weighted images (DWIs) required. Diffusion tensor imaging (DTI)[28–30] needing at least six DWIs and one non-diffusion-weighted image. To accurately estimate fiber crossings and the properties of intra- and extra-cellular compartments, advanced methods[31–38] typically require more DWIs. While this increases data volume, it presents an exciting opportunity to enhance the precision of dMRI. Efforts are ongoing to streamline acquisition and reconstruction processes.

Artificial intelligence (AI) empowered methods have emerged as powerful tools for accelerating acquisition[39] and reconstruction[40–44], as well as multi-parametric estimation[45,46]. AI-based joint reconstruction techniques leverage correlations across k-space, spatial-space, and q-space to allow for sparser sampling while compensating for missing data[47–49]. However, most of these methods rely on simulated k-space data synthesized from already-reconstructed magnitude images, rather than empirically acquired k-space data. The use of empirical raw k-space dMRI data is vital for training more robust and accurate AI models.

In this paper, we present Diff5T Dataset, the first 5.0T brain imaging dataset containing both k-space and image-space data. This dataset includes high-resolution T1-weighted (T1w) and T2-weighted (T2w) imaging, along with multi-shell, multi-direction, and multi-channel dMRI data from 50 subjects. Additionally, we provide fully documented reconstruction and preprocessing pipelines to support reproducible and further research. Diff5T represents a significant step forward in addressing the challenges of high-resolution dMRI, offering a critical resource for the imaging and neuroscientific communities.

## Methods

### Participants.

This study was approved by the Institutional Review Board/Ethics Committees of Shenzhen Institutes of Advanced Technology, Chinese Academy of Sciences, with a requirement for written informed consent. All patient records were de-identified before analysis and were reviewed by the institutional review boards to guarantee no potential risk to patients.

As part of the written consent process, all data were made publicly accessible. Participants were thoroughly informed about the significance of the study and provided written consent for the future public release of their data. The inclusion criteria for participation were as follows: (1) healthy adults with no history or diagnosis of brain disease, and (2) availability for an MRI examination encompassing all required imaging sequences.

A total of 50 healthy volunteers, aged 18–38 years, were recruited for the study, each providing written informed consent in accordance with ethical guidelines. These data have been subjected to quality control by data collectors and physicians with more than 20 years of extensive clinical experience.

## Data acquisition.
### Hardware
Data were acquired using a 5.0T MRI scanner (uMR Jupiter, United Imaging Healthcare, UIH, China) equipped with a maximum gradient strength of 120 mT/m and a slew rate of 200 T/m/s. Imaging was performed with a quadrature birdcage transmit coil and a 48-channel receiver coil (48/2 channel Rx/Tx head coil)[27]. To maintain consistency across imaging sessions, field shim settings were standardized to reduce image distortion differences. During the scans, three sponge pads were placed around the participants' heads to minimize motion artifacts and ensure stability throughout the imaging session. The movement of patients was also recorded during the scans.

### Protocols
A standard scan session order is as follows: localization, three dMRI series, T1w and T2w series, as shown in Fig. 1.

### dMRI data
To achieve 1.2-mm isotropic spatial resolution dMRI data with a balanced trade-off between SNR efficiency and scan time, the echo planer imaging (EPI) sequence was used in dMRI acquisition with the following parameters: TR/TE = 8277/67.9 ms; Echo spacing = 0.8 ms; Flip angle = 90°; In-plane FOV (RO x PE) = 212 × 212 mm$^2$; Number of slice = 114; Bandwidth = 1850 Hz/Pixel; Multi-band acceleration factor = 2; In-plane acceleration factor = 2; Partial Fourier factor = 6/8; Acquisition matrix (RO x PE) = 176x 176; Total acquisition time = 45.2 minutes. The dMRI images were obtained in the axial orientation.

To achieve high angular resolution, 291 volumes of dMRI data were acquired in a standard scan session, consisting of 90 DWIs at b=1000 s/mm$^2$, 90 DWIs at b=2000 s/mm$^2$ DWIs and 90 DWIs at b=3000 s/mm$^2$ DWIs. The phase encoding was along posterior-to-anterior (PA) direction. A total of 21 b=0 image volumes were acquired, with 6 of them acquired along AP phase encoding direction and 15 of them acquired along PA phase encoding direction.

The uniform multi-shell diffusion gradient vectors were optimized by the spherical code method[50] in DMRITool, which directly maximize the minimal separation angles between different diffusion samples from both within and across shells. More detailed imaging parameters are listed in Table 1.

### Structural data
Two high resolution structural series (i.e., T1w and T2w) were also acquired following the dMRI series in each session.

The 0.5-mm isotropic T1w images were acquired with the GRE_FSP (Fast Spoiled Gradient Echo) sequence with the following parameters: TR/TE = 10.1/3.4 ms; TI = 1000 ms; ETL = 165; Flip angle = 9°; FOV = 256 × 256 × 150 mm$^3$; Bandwidth = 160 Hz/Pixel; uCS (United Imaging Compressed Sensing) combined acceleration factor = 3; Total acquisition time = 13.6 minutes. The T1w images were obtained in the sagittal orientation.

The 0.5-mm isotropic T2w images were acquired with the FSE_MX3D (Fast Spin Echo Modulated flip Angle Technique in Refocused Imaging with eXtended echo train) sequence with the following parameters: TR/TE = 3000/421.68 ms; ETL = 200; Flip angle mode = T2; Flip angle (max/min) = 160/23°; Reference tissue T1/T2 = 1800/96 ms; FOV = 256 × 212 × 150 mm$^3$; Bandwidth = 450 Hz/Pixel; uCS combined acceleration factor = 3; Total acquisition time = 13.7 minutes. The T2w images were obtained in the sagittal orientation.

## Data reconstruction.

The general pipeline to produce the Diff5T dataset is illustrated in Fig. 1.

The online reconstructed DICOM images were converted to compressed NIfTI format using *dcm2niix*[51] tool. The online reconstructed data are hereafter referred to as 'DICOM-to-NIfTI' data.

The offline reconstructed NIfTI images retain the same basic information as the DICOM-to-NIfTI images, such as the affine transformation matrix. The offline reconstructed data are hereafter referred to as 'reconstructed' data. The details of offline reconstruction are as follows:

### dMRI data

The raw files (.raw) were directly exported from the 5T scanner. Using the offline UIH toolbox (UIHRawdata2ISMRMRD) written in MATLAB (MathWorks, Natick, MA, USA), the k-space data were extracted from the raw files and saved in binary format (.bin). This process ensured the removal of all personal identifiers, including subject name, birthdate, height, and weight. The following steps were used to perform offline reconstruction of dMRI k-space data in MATLAB.

- **Pre-whitening**

Pre-scan noise data were used to pre-whiten the k-space data, which helped remove complex correlations and structures in its statistical properties. This process made the noise more predictable and easier to model[52,53].

- **Ghost correction**

Nyquist Ghost Corrected (NGC) or Ghost Elimination via Spatial and Temporal Encoding (GESTE)[54] was used to suppress Nyquist ghosts by a fusion of reported temporal[55] and spatial[56] encoding methods. Generally, GeneRalized Partially Parallel Acquisitions (GRAPPA) methods estimate missing points in accelerated measurement data using kernels trained from auto-calibration signal (ACS) data. However, different sampling patterns and readout polarity shifts due to eddy currents cause kernel estimation errors, leading to ghosting artifacts[57]. Thus, these two methods were used to adjust the reference and accelerated k-space data to match the same pattern. Empirically, NGC was used to correct raw k-space data and muti-band ACS data, while GESTE was used to correct in-plane ACS data.

- **Inter-slice unaliasing**

Slice-GRAPPA[58] was used to disassemble the slice-aliased images and reconstruct multi-band accelerations in parallel imaging. Slice-GRAPPA has been shown to significantly reduce image artifacts in *in-vivo* acquisitions[58].

- **In-plane unaliasing**

1D GRAPPA[59] was used to unaliased the in-plane data by synthesizing the missing data points directly in k-space. To estimate the full k-space data, Projection Onto Convex Set (POCS)[60] was implemented for Partial Fourier reconstruction to recover the k-space data by utilizing the conjugate symmetry of k-space.

- **Muti-channel combination**

To obtain the complex-valued images from multi-channel data, an adaptive combination method[61] was implemented to combine the multi-channel images with lower noise floor than sum-of-squares method[62].

- **PCA denoising**

The eigen spectrum of random covariance matrices of dMRI data can be priorly described by a Marchenko-Pastur distribution[63,64]. Moeller et al.[65,66] proposed the NOise Reduction with DIstribution Corrected (NORDIC) method as a noise reduction framework for diffusion MRI, which leverages low-rank modelling of gfactor-corrected complex dMRI reconstruction and non-asymptotic random matrix distributions to remove noise signal. Finally, the magnitude is taken as the reconstructed image.

After these steps, the reconstructed complex-valued k-space data and the reconstructed magnitude-valued dMRI images were obtained.

### Structural data

The reconstruction of the T1w and T2w images were performed using vendor's commercial UIH online reconstruction software (including vendor's implementation of parallel imaging reconstruction, fast Fourier transformation, coil combination, root mean square combination of different echo images and calculation of field maps) and were exported as DICOM format from scanner.

### Data preprocessing.

The following preprocessing steps were applied to both the DICOM-to-NIfTI images and the reconstructed images.

### dMRI data

The preprocessing pipeline for dMRI data is shown in Fig. 1. Three dMRI series were collected, each corresponding to a distinct q-space shell, with five b0 images included in each series. After preprocessing each series individually, all three were concatenated in their original scanning order as the final step. The preprocessed DICOM-to-NIfTI images are shown in Fig. 3a. described in detail as below:

- **Degibbs**

  Gibbs ringing appears in MRI images as spurious oscillations around sharp tissue boundaries, resulting from the truncation of k-space sampling. Depending on the location of the sharp edge relative to the sampling grid, this can lead to attenuation of the Gibbs ringing artifact. The used method *mrdegibbs* in MRtrix3 toolbox (version 3.0.4)[67] tries to estimate the subvoxel-shift in pixels necessary to minimize this distance and interpolates the image accordingly[68].

- **Distortion correction**

  The susceptibility-induced and the eddy current-induced off-resonance field distortions were estimated and corrected using *topup*[69,70] and *eddy*[71–73] (GPU version *eddy_cuda10.2*) methods in FSL toolbox (version 6.0.7.14)[70,74]. The field maps were estimated from one b=0 image pairs acquired with reversed PE blips using *topup* method. The *topup* outputs were used to perform the *eddy* method. The *eddy* method effectively corrected for movement of subject, eddy current-induced and susceptibility-induced distortion. The brain mask for *eddy* correction was generated from the mean b = 0 image volume after *topup* correction, using *bet2*[75] with a fractional intensity threshold of 0.12. To account for head motion, *eddy* adjusted the diffusion directions, leading to small variations for each participant.

- **Filed bias correction**

  The N4 bias field correction algorithm is a popular method for correcting low frequency intensity non-uniformity present in MRI image data. The bias field, caused by magnetic field inhomogeneity during acquisition, negatively impacts any intensity-based segmentation. Thus, the N4 bias field correction algorithm in ANTs was implemented by using MRtrix3 command *dwibiascorrect ants*[67,76].

- **Registration to structural**

  Finally, the undistorted mean b0 image obtained from *topup* was registered to the T1w structural image using a boundary-based registration (BBR)[77]. The dMRI data (b ≠ 0) were then transformed according to structural volume space by the estimated displacement field.

### Structural data

Refer to HCP minimal pipelines[78], the 0.5-mm isotropic T1w and T2w volumes were aligned to the standard MNI space template (0.7-mm isotropic) with rigid body registration. Then, cerebral cortical surface reconstruction and volumetric segmentation were performed using the "recon-all" function of the FreeSurfer software (version v6.0.0)[79–81]. Finally, T2w were registered to T1w using BBR to ensure alignment across tissue boundaries. Facial features were obscured in each image volume by face masking algorithm[82] in XNAT-Tools.

### Microstructural modelling.

To demonstrate the range of analyses possible with this comprehensive dataset, several diffusion models were applied to estimate the microstructural complexity of axons *in vivo*. These models, commonly used in clinical and neuroscience research, are supported by publicly available software and codes.

- **DTI**

  To estimate the fiber orientation, the diffusion tensor imaging (DTI)[28–30] was fitted to the b = 1000 s/mm2 data using the *dtifit* method from FSL.

- **NODDI**

  The neurite orientation dispersion and density imaging (NODDI)[36] parameters were estimated using Accelerated Microstructure Imaging via Convex Optimization (AMICO)[83] method on the b = 1000, 2000, 3000 s/mm$^2$ data.

- **MSMT-CSD**

  The multi-shell multi-tissue constrained spherical deconvolution (MSMT-CSD)[84,85] was conducted on the b = 1000, 2000, 3000 s/mm$^2$ data using the *dwi2response dhollander* method. Fiber Orientation Distribution (FOD) maps for white matter, grey matter, and cerebrospinal fluid, represented in the Spherical Harmonic (SH) basis, were obtained by *dwi2fod msmt_csd*. The *fod2dec* method was used to generate the FOD-based DEC, which was scaled according to the integral of the FOD. The above methods are implemented using the MRtrix3 tool.

- **Tractography**

  Streamlines tractography was performed using the *tckgen* method with the probabilistic Second-order Integration over Fiber Orientation Distributions (*iFOD2*) algorithm, which integrates second-order FODs. The white matter FOD map, derived from *dwi2fod*, was used as the input to *tckgen*. The "-minlength" and the "-maxlength" were set to 40 and 200 respectively. The desired number of streamlines "-select" was set to 10 million. The above methods are implemented using the MRtrix3 tool.

## Data Records

The four types of data records listed in this section are publicly available in Diff5T dataset. The k-space data are in binary MAT-format (.mat) and the image data are in compressed NIfTI format (.nii.gz). The T1w and T2w data are only included in the DICOM-to-NIfTI images, while the dMRI data are included in all types of data.

The un-preprocessed and preprocessed image data were released for researchers to explore further preprocessing methods. The offline reconstruction and preprocessing scripts are also available.

### Raw k-space data.

The data include raw k-space data, multi-band and in-plane navigation k-space data for ghost correction, as well as multi-band and in-plane reference k-space data as the ACS data for reconstruction.

### Reconstructed k-space data.

The data include processed k-space data. After performing pre-whitening, ghost correction, and estimation of undersampled k-space (including slice-GRAPPA, in-plane GRAPPA and POCS) on the raw k-space as shown in Fig. 1b, the reconstructed multi-channel k-space data was obtained.

### Reconstructed images.

The reconstructed images were obtained from the reconstructed k-space data using Fourier transformation, adaptive multi-channel combination, and other preprocessing steps as shown in Fig. 1.

### DICOM-to-NIfTI images.

The DICOM data consists of spatially resolved images, with the raw data discarded during the acquisition process. DICOM files contain a greater variety of scanning settings than the raw k-space data. To standardize the image format, DICOM files were converted to NIfTI format. This conversion process anonymized the images while retaining essential acquisition parameters in .json files, which include the field of view (FOV), acquisition matrix, number of slices, slice thickness, number of coils, TR/TE, flip angle, down-sampling factor, and other relevant details.

## Technical Validation

### Assessment of diffusion data quality.

To assess the feasibility of using the provided k-space data for image reconstruction task, a series parallel reconstruction methods (Fig. 1b) were employed as benchmark examples. The reconstructed images for all participants were visually inspected for quality control. Fig. 2 shows the DICOM-to-NIfTI images on the left, and the reconstructed images using the pipeline in Fig. 1b on the right. In comparison, the signal-to-noise ratio (SNR) of DICOM-to-NIfTI images is relatively higher, potentially because vendor's pipeline applied some sort of denoising. The reconstructed images exhibited slightly higher background noise levels compared to DICOM-to-NIfTI images. However, the reconstructed images appeared smoother, with clearer edge details, particularly for b=3000 images. It is worthy to note that the pipeline adopted in this work is not optimal. Researchers are encouraged to refine this pipeline or employ their own to achieve higher-quality results.

### Assessment of participant head motion.

Participant head motion during the scan was evaluated, as shown in Fig. 3b.
This motion was quantified in each shell using FSL eddy, specifically using restricted movement root mean squared (RMS) outputs[71] which show RMS movement in each volume relative to the previous volume.

### Assessment of microstructural modelling results.

The quality of the diffusion MRI data is also assessed by evaluating the microstructural modelling results using different methods, shown in Fig. 4.

Different microstructural metrics were derived from two distinct diffusion-based models of DTI and NODDI. Specifically, maps of mean diffusivity (MD), fractional anisotropy (FA) and directionally encoded colour (DEC) FA maps are displayed for DTI, as shown in Fig. 4a. Maps of intra-cellular volume fraction (Vic), isotropic cerebrospinal fluid volume fraction (Viso), and orientation dispersion (OD) are displayed for NODDI, as shown in Fig. 4b.

In terms of fiber orientation distribution estimation and tractography, the fiber orientation distribution (FOD) was fitted using the MSMT-CSD algorithm and the whole brain tractography was fitted using *iFOD2* algorithm. Fig. 4c shows the volume fraction maps of white matter (WM-VF) and gray matter (GM-VF), as well as the DEC FA maps weighted by the integral of the FOD. Fig. 4d shows FOD and streamlines tractography within the left centrum semiovale and the coronal radiata of a representative participant. The results in these two regions are visually consistent with the anatomical structure.

These methods have different data requirements for estimation, which are all met by this dataset. The resulting estimates are consistent with those reported in the relevant literature and visually align with human brain's anatomy.

## Code Availability

The MATLAB code for the dMRI reconstruction described above are included in the released dataset. The HCPpipelines and the Connectome Workbench for co-registration are available publicly available (https://github.com/Washington-University/HCPpipelines; https://github.com/Washington-University/workbench). The FMRIB Software Library used in the diffusion data pre-processing and DTI model fitting is publicly available (https://fsl.fmrib.ox.ac.uk/fsl/docs/#/install/index). The MRtrix3 software for the diffusion data pre-processing, constrained spherical deconvolution and tractography generation is publicly available (https://www.mrtrix.org/). The Advanced Normalization Tools (https://github.com/ANTsX/ANTs). The software used for face masking is publicly available (https://wiki.xnat.org/xnat-tools/face-masking). The FreeSurfer software is publicly available (https://surfer.nmr.mgh.harvard.edu/fswiki/DownloadAndInstall). The Accelerated Microstructure Imaging via Convex Optimization for fitting NODDI model is publicly available (https://github.com/daducci/AMICO). The DMRITool for optimizing uniform multi-shell gradient vectors is publicly available (https://diffusionmritool.github.io/).

## Usage Notes

### Data hosting.

The dataset is publicly accessible, and the data along with the processing code will be gradually uploaded and updated in the coming period. The example data and processing code can be accessed at the GitHub repository (https://github.com/ShoujunYu/Diff5T).

## Acknowledgements

This research was partly supported by the National Key Research and Development Program of China (No. 2022YFA1004200), the National Natural Science Foundation of China (No. 52293425, No. 62222118, No. U22A2040), Shenzhen Science and Technology Program (No. RCYX20210706092104034, No. JCYJ20220531100213029), Shenzhen Medical Research Fund (No. B2402047), Key Laboratory for Magnetic Resonance and Multimodality Imaging of Guangdong Province (No. 2023B1212060052), and Youth lnnovation Promotion Association CAS, the Beijing Natural Science Foundation (Grant No. L242038), the Open Research Fund of the State Key Laboratory of Brain-Machine Intelligence，Zhejiang University (Grant No. BMI2400001)

## Author contributions

S.W. and S.Y. proposed the idea and initialized the project and the collaboration.
S.W. and S.Y. developed and implemented the framework.
S.W., Y.X., J.C., C.T. and Y.D. provided the protocols.
S.Y. and S.W. collected the brain MRI data.
S.Y., S.W., S.J., J.Z. and H.P. developed the reconstruction workflow.
S.Y., S.W., J.C., S.J., J.H., F.Z., Q.T., H.G. and G.L. analysed the results and plotted the figures.
S.Y., S.W. and J.C. wrote the manuscript.
S.W. and H.Z supervised the project.
All authors read and contributed to revision and approved the manuscript.

## Competing interests

The authors declare no competing interests.

## Figures

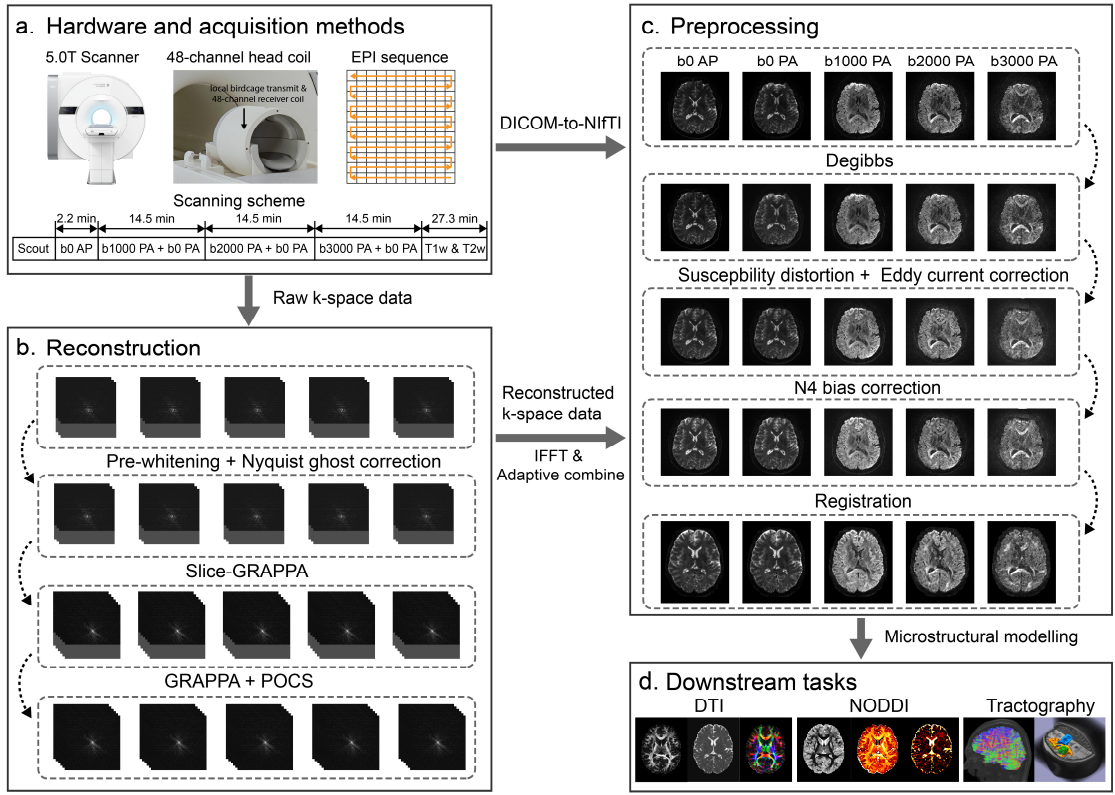

Fig. 1 - Diff5T data processing pipelines.

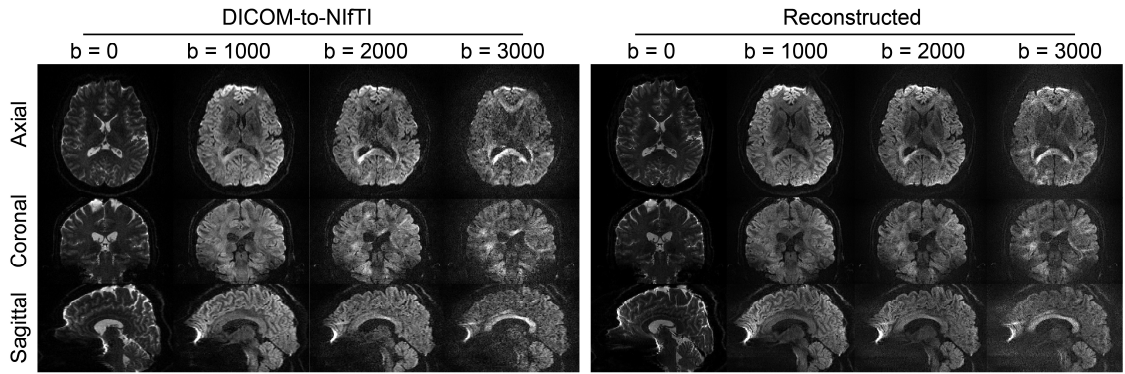

Fig. 2 - dMRI reconstruction.

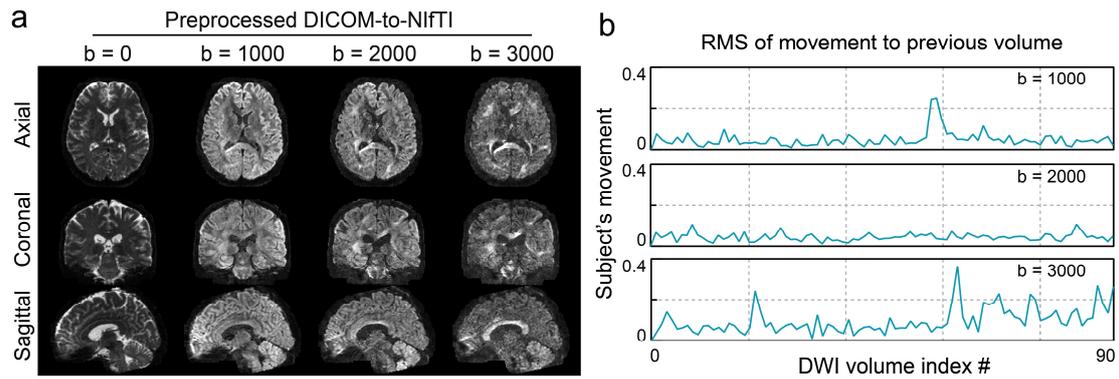

Fig.3 - dMRI preprocessed.

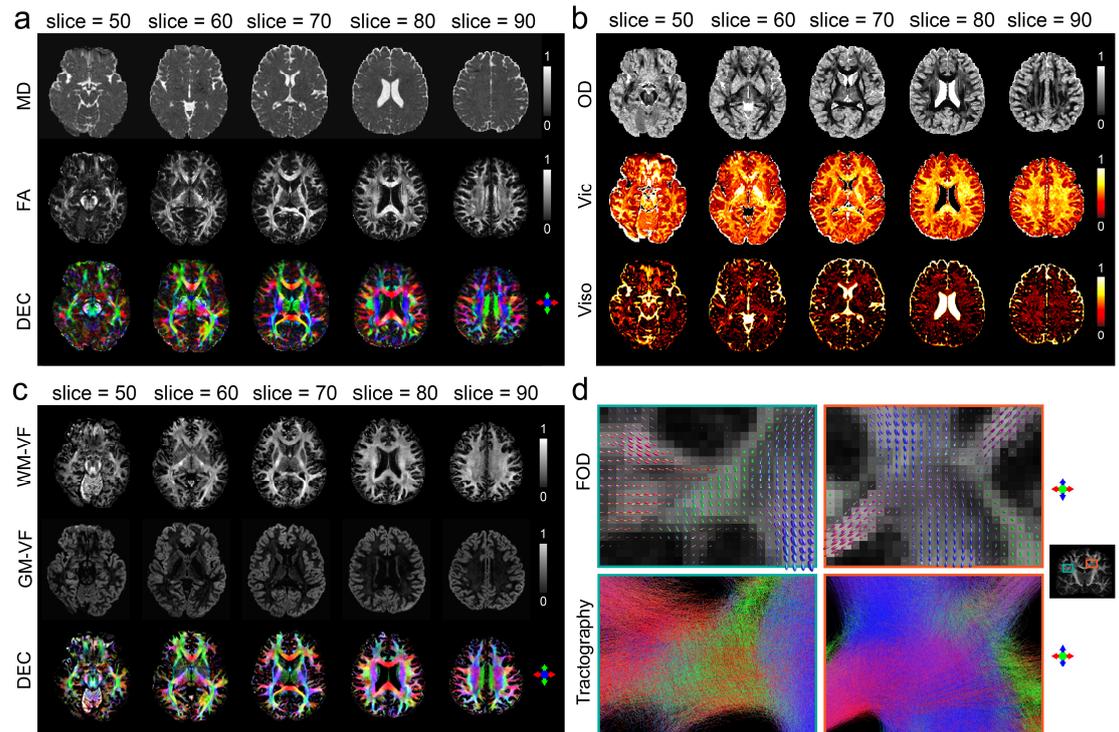

Fig. 4 - Microstructural modelling.

# Figure Legends

**Fig. 1 - Diff5T data processing pipelines.**

a) The hardware and acquisition methods used to acquire the data. The Jupiter 5.0 T scanner, the 48-channel head coil, the EPI acquisition technique and the acquisition protocols. b) The reconstruction pipeline. The dMRI data were pre-whitened and corrected for Nyquist ghosts, then reconstructed using slice-GRAPPA, GRAPPA, and partial Fourier reconstruction methods. The structural images were reconstructed on the scanner. c) dMRI preprocessing pipeline. The preprocessed dMRI data were obtained following Gibbs ringing removal, susceptibility- and eddy current-induced distortion correction, bias field correction, and final registration to resampled T1-weighted image. d) dMRI downstream tasks. Fitting the dMRI data to DTI and NODDI models and performing tractography.

### Fig. 2 - dMRI reconstruction.

The left diffusion weighted images are DICOM-to-NIfTI conversions obtained through online reconstruction. The right diffusion weighted images are obtained through offline reconstruction.

### Fig. 3 - Preprocessed dMRI.

a) The preprocessed DICOM-to-NIfTI images were obtained following the preprocessing steps illustrated in Fig. 1c. b) The root mean squared (RMS) voxel-wise displacement relative to the previous diffusion weighted volume, calculated using FSL's *eddy*.

### Fig. 4 - Microstructural modelling.

a) The MD, FA, and DEC derived from the tensor were fitted to a single-shell (b=1000 s/mm2) DTI model using FSL's *dtifit*. b) The OD, Vic, and Viso parameters were obtained by fitting a multi-shell (b=1000, 2000, 3000 s/mm2) NODDI model using the AMICO toolbox. c) The WM-VF, GM-VF, DEC were estimated by using the MRTrix3 toolbox. d) The FOD and Tractography of the left centrum semiovale and the coronal radiata.

## Tables

### Table 1. The Information on Data Acquisition.

| Parameters | dMRI | T1-weigted | T2-weigted |
|---|---|---|---|
| Sequence | EPI | GRE-FSP | FSE-MX3D |
| Acquisition orientation | Axial | Sagittal | Sagittal |
| TR / TE (ms) | 8277 / 67.9 | 10.1/3.4 | 3000/421.68 |
| FA (degree) | 90 | 9 | Max = 160, Min = 23 |
| Echo spacing (ms) | 0.8 | / | 5.02 |
| ETL | 66 | 165 | 180 |
| Band width (Hz/Pixel) | 1850 | 160 | 450 |
| In-plane resolution (RO × PE mm$^2$) | 1.2 × 1.2 | 0.5 × 0.5 | 0.5 × 0.5 |
| Acquisition matrix (RO × PE) | 176 × 176 | 512 × 512 | 512 × 424 |
| Slice thickness (mm) | 1.2 | 0.5 | 0.5 |
| Number of slices | 114 | 300 | 300 |
| Multi-band factor | 2 | / | / |
| In-plane acceleration factor | 2 | / | / |
| Partial Fourier | 6/8 | / | / |
| Acceleration method | GRAPPA | uCS | uCS |
| Combined Acceleration factor | 4 | 3 | 3 |
| Channel combine mode | Adaptive combine | Adaptive combine | Adaptive combine |
| Fat suppression | Fat Sat | Off | Off |
| Intensity uniformity correction | Off | Combined field | Combined field |
| Distortion correction | Off | 2D | 2D |
| Image Filtering | Off | Smoothing: 5 Enhancement: 1 Edge Smoothing: 4 | Smoothing: 5 Enhancement: 3 Edge Smoothing: 3 |
| k-space filtering (mode/intensity) | Standard / High | Default / Medium | Optimized / Medium |
| Other | b = 0, 21 volumes (6 AP, 15 PA) b = 1000 s/mm$^2$, 90 directions (PA) b = 2000 s/mm$^2$, 90 directions (PA) b = 3000 s/mm$^2$, 90 directions (PA) | TI = 1000 ms | |
| Scanning time | 45.2 min | 13.6 min | 13.7 min |

Notes. EPI = echo planer imaging; GRE-FSP = fast spoiled gradient echo; FSE-MX3D = fast spin echo-modulated flip angle technique in refocused Imaging with extended echo train; TR = repetition time; TE = echo time; TI = inversion time; FA = flip angle; ETL = echo train length; RO = read out; PE = phase encoding; uCS = United Imaging compressed sensing.